# SS 433: flares in H$\alpha$, GRAVITY observations & $L_2$ ejection

M. G. Bowler

Department of Physics, University of Oxford, Keble Road, Oxford OX1 3RH, UK
e-mail: michael.bowler@physics.ox.ac.uk

## Abstract

The microquasar SS 433 exhibits in H$\alpha$ intermittent flares, Doppler shifted to both the red and the blue. The mean remembers the orbital phase of the compact object. I show that the flares are not intermittent sightings of an accretion disk; rather, plasma must be expelled through the $L_2$ point, thus remembering the phase of the orbit as it invades the space beyond the system. That space has been mapped with GRAVITY observations of a similar flare, revealing a strong rotation component.

## Introduction

A sequence of spectra of SS 433 taken in 2004 showed the onset of an optical flare, with daily variations characteristic of H$\alpha$ radiation from a nearly edge on accretion disk, the rim orbiting at speeds $\sim$700 $km\ s^{-1}$. The radius of the corresponding orbit would have to lie deep within the photosphere surrounding the compact object. Analysis of the continuum background, hitherto largely ignored, reveals no evidence of changes in the photosphere with onset of flaring. The H$\alpha$ flares seem consistent with a very different origin, episodic overflow of material from the Companion through the $L_2$ point. This may be relevant to GRAVITY observations in Br$\gamma$ of material far outside the system, with rotational speeds of up to $\sim$1000 $km\ s^{-1}$.

Among early observations of the microquasar SS 433 with the GRAVITY instrument, which takes spectra in the near infrared as a function of angular position on the sky, is an episode revealing in Br$\gamma$ material with rotational speeds of several hundred $km\ s^{-1}$, more or less in the plane of the orbit of the binary and at radii of several times the binary separation $A$, Fig.3 of [1].These observations are at least superficially consistent with a circumbinary disk, but with supposedly orbital speeds of $\sim$100 $km\ s^{-1}$ at radii of $\sim$ 10 $A$ and $\sim$ 1000 $km\ s^{-1}$ at $\sim A$ the contained mass would have to be $\sim$ 400 $M_{\odot}$ for the specific angular momentum of $\sim$ 1000 $A\ km^2 s^{-1}$. However, the circumbinary disk of SS 433 orbits the system with a speed of $\sim$ 240 $km\ s^{-1}$ at a radius of < 2 $A$ [2] and is long lasting and very stable; the interior mass is $\sim$ 40 $M_{\odot}$. [2,3]. The authors of [1] suggest that their observations are of super-Kepler material, either ejection of the circumbinary ring or perhaps from eruptions in the core of the accreting material. The high specific angular momentum remains a problem. Material leaking out of the Roche lobe of the compact object through the $L_2$ point would share its specific angular momentum, a mere 375 $A$. Any additional gravitational torquing seems unlikely to increase this to more than $\sim$500$A$. It is therefore natural to ask if there are other data looking superficially like a disk rotating at speeds of up to $\sim$ 1000 km $s^{-1}$. There are indeed.

**Optical flares in H$\alpha$**

Optical flares occur in SS 433 at irregular intervals and for irregular periods of time. A characteristic feature is the appearance of spectral line splitting by as much as $\pm 1000\ km\ s^{-1}$, [4], which may last for a number of days [5,6; see also 7]. This phenomenon has been interpreted as glimpses of the outer regions of the accretion disk [5]. Should this interpretation be wrong, the phenomenon must be attributed to bursts of a wind or mass ejections of some kind. Any origin in the close vicinity of the compact object cannot account for the fast rotating and radially distant source of the GRAVITY Br$\gamma$ observations.

Given the data of [1], I have re-examined the supposed accretion disk data, primarily those of H$\alpha$ in [5], looking for any indication so far not considered that might fix for these data an origin either in an accretion disk or alternatively in ejected super-Kepler material, not bound within the system. I discuss the evidence for and against each of these interpretations. Accreting material moving at $\gtrsim 700\ km\ s^{-1}$ in Kepler orbits must be within $0.1A$ of the compact object; it emerges that it would be too close in to be visible. An origin in material ejected from the system via the $L_2$ point now seems much more probable.

The optical spectra of SS 433 contain emission lines radiated by the relativistic jets and lines radiated from matter moving much more slowly within and close to the system; the stationary lines. He II 4685 Å traces the orbital motion of the compact object through a component that is eclipsed by the Companion; a pair of C II emission lines are likewise eclipsed [8]. The vast majority of emission lines are not eclipsed, being formed outside of the system in the circumbinary disk or above and below the orbital plane in polar winds. Predominant are hydrogen lines H$\alpha$ and the He I lines at 6678, 7065 Å [9]; similar phenomena are encountered in the infrared [7,10]. Lines radiated from the circumbinary disk move, if at all, only as a result of the relative intensities of the red and blue shifted components oscillating. All these features are clear when SS 433 is optically quiescent, much less so during an optical flare. The principal feature of an optical flare is the appearance of red and blue shifted components (particularly studied in H$\alpha$) separated by Doppler speeds of between 1000 and 2000 $km\ s^{-1}$. A striking example is to be found in Fig.1, bottom left panel, of [4].

The most systematically complete set of data is that of [9], primarily in H$\alpha$ and He I emission lines. The stationary lines were quiescent between Julian day 2453000 +245 and +287; analysis revealed the circumbinary disk and the polar winds from the environs of the compact object are sufficient to account for the spectra. From +287 the flare commenced and the stationary lines became much broader; as the disk turned more edge on, absorption troughs occurred. Over the period of observation the separation of the red and blue flare components in H$\alpha$ increased from $\sim$ 1000 to $\sim$ 1400 $km\ s^{-1}$. These red and blue components (see Fig. 1) together traced out the motion of the compact object with some precision over the first orbital half period, Fig.2.

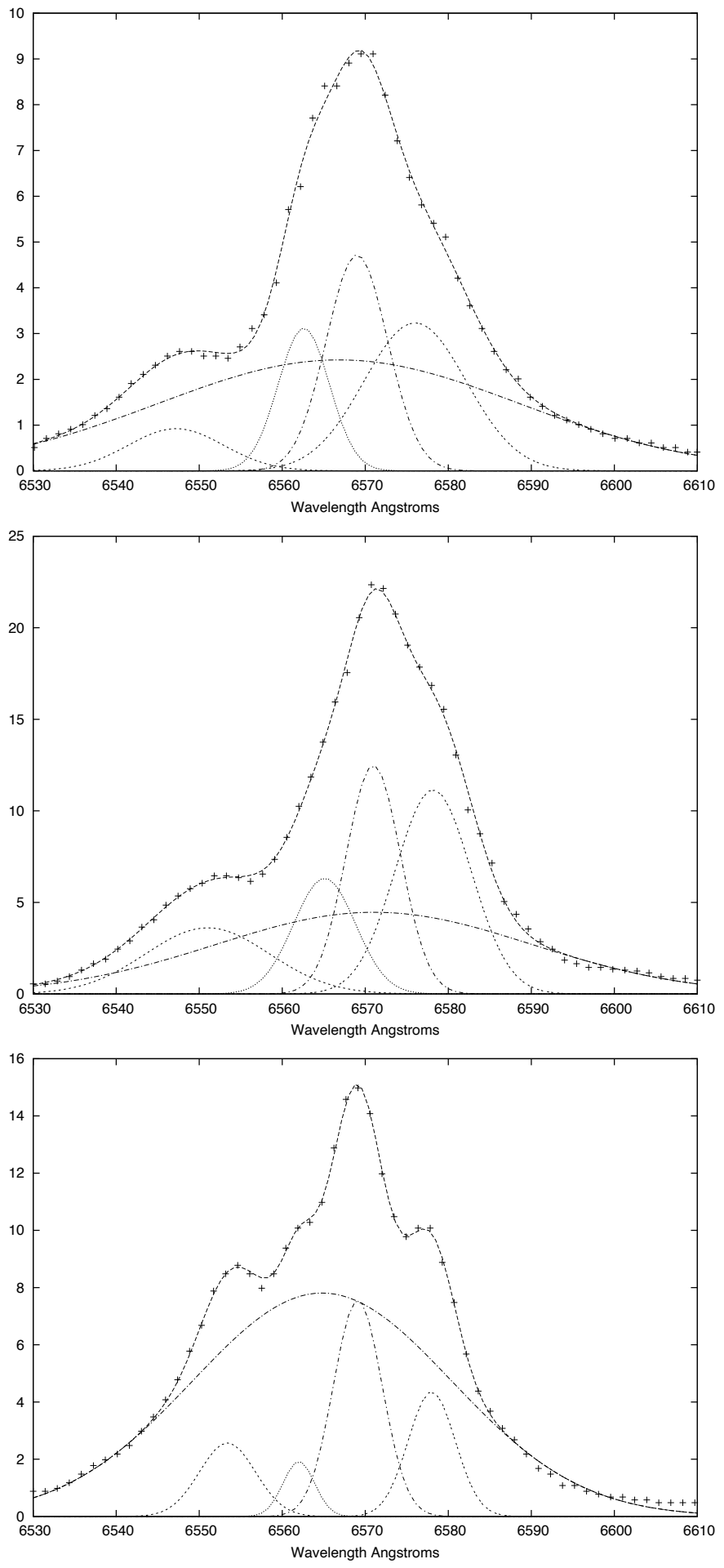

Fig.1 Three spectra of H$\alpha$ taken from [5], after the onset of flaring. The *lowest panel* is for day +291, the *middle* for +294 and the *uppermost* +297. Each is decomposed into five gaussian components; the two outermost are flare. Day +294 was an eclipse day; both the blue and the red flares are substantial, having roughly twice the normalised intensity of the other two examples.

The relationship became less precise as time evolved but retained some memory; the He I data are confounded by the moving jet lines during this latter period. The obvious interpretation was that these H$\alpha$ flare components constitute glimpses of the accretion disk of the compact object, appearing as a result of a burst of accretion or, if accretion be continuous, of a parting of the clouds. The details are given in [5] and the observations placed in a wider context in [6]. It is of course the way in which the widely separated H$\alpha$ components swing together, following or remembering the motion of the compact object, that is the strongest evidence suggestive of an accretion disk rim orbiting at $\approx 700\ km\ s^{-1}$.

After flareup, the compact object and its environs are eclipsed by the Companion on days + 294 and + 307; on neither occasion are either of the extreme components of H$\alpha$ eclipsed (Figs. 1,2,4). If these components originate from material in a Kepler orbit about the compact object, of mass $\sim 15\ M_{\odot}$ , the radius of that orbit must be $\sim 0.07A$.

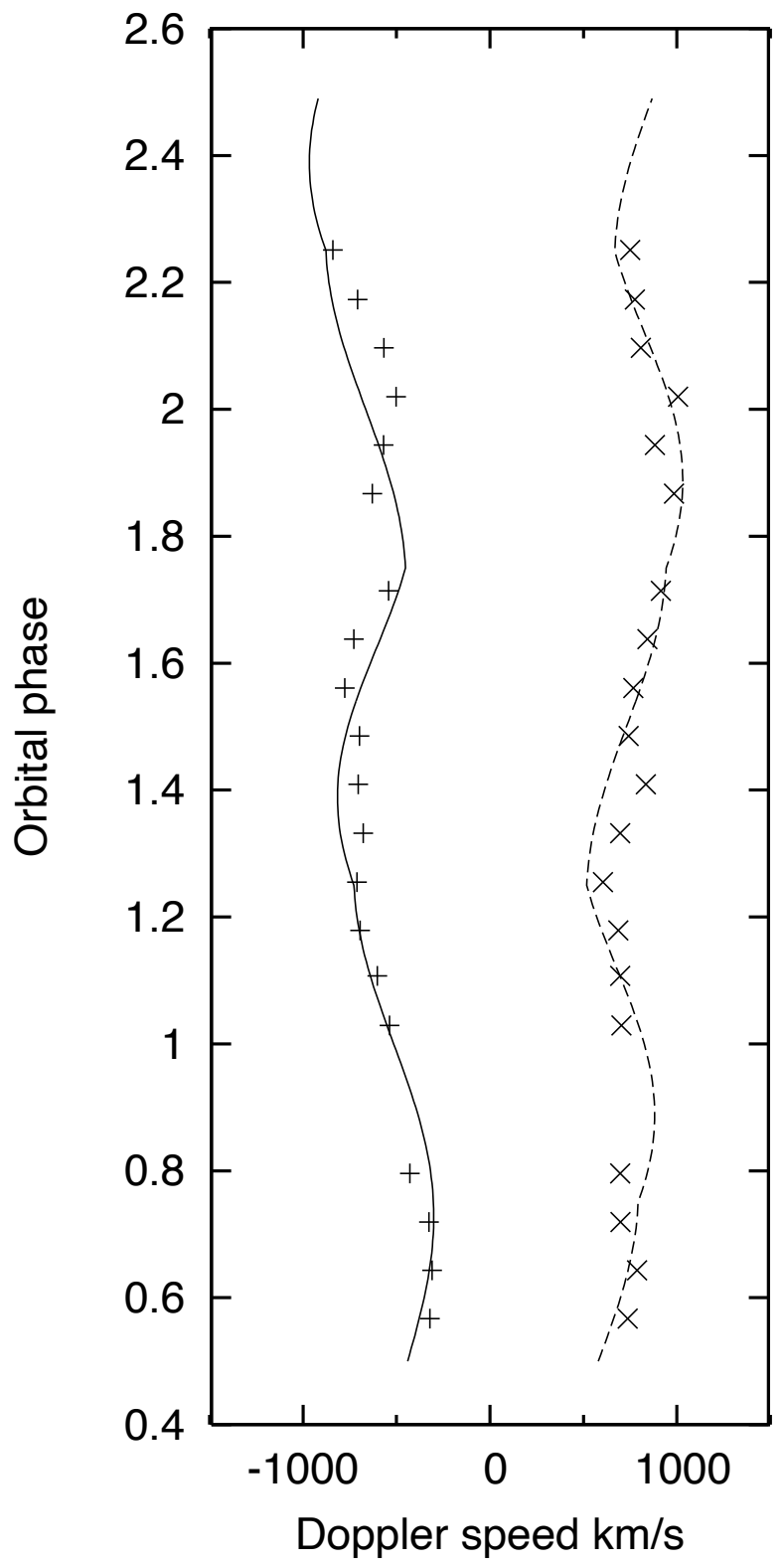

Fig. 2 The Doppler shifts of the blue and the red flare components as orbital phase advances (from [5]). This pattern reveals a distinct memory of orbital phase. The superimposed curves assume a source in an accretion disk; the memory could also be caused by $L_2$ expulsion of plasma (see text).

The absence of eclipses of such material would imply that the radius of the companion is $a< 0.25A$ [5], provided the orbit is adhering to the Goranskii ephemeris [11]. The radius of the Roche lobe of the donor is $\sim 0.4A$ and it is expected that the donor will fill or over fill this lobe, with stable transfer of material. In order to search for further evidence for or against an accretion disk origin, it is relevant to search the data of [9] for indications of how well the system followed the Goranskii ephemeris and how readily eclipses of these H$\alpha$ components could be avoided if the Companion does in fact fill its Roche lobe.

Stimulated by the spatially resolved data obtained by the GRAVITY observations in [1] I have re-examined the data of [9] which I previously analysed in [5]. In the first instance I concentrated on the absence of any eclipses of the red and blue flare components. The data of [9] cover eclipses of the compact object at approximately days +254, 268, 281 then, after the flare, at 294 and 307. The orbital phases (according to the Goranskii ephemeris) of the spectra closest to eclipse are respectively 0.97, 0.04, 0.035, 0.03 and 0.02. Days +294 and 307 are well after the onset of the flaring activity.

The eclipse timetable is based on photometry and it is well known that in the V band the eclipse lasts over 2 days and at minimum the partially eclipsed signal has dropped by a factor of about 2. This suggests that the photosphere surrounding the compact object, not completely eclipsed, must have a similar radius to that of the (eclipsing) Companion and that these radii are $\sim A/2$. If the system does not deviate from the Goranskii ephemeris the absence of eclipses of the red and blue features at about 700 $km\ s^{-1}$ suggests that the Companion radius $a$ is < 0.25 $A$ [5]. It is therefore of some significance to check on the extent to which the system deviates from that ephemeris. The data of [9] are not photometric, but not devoid of relevant information. The line spectra have been normalised to the local continuum and so any line that is relatively stable and not eclipsed doubles in (normalised) height as the compact object passes through eclipse. Up to day +281 most spectral lines are formed in jets, the circumbinary disk or the winds above the accretion disk and show no sign of the disturbances accompanying the later flaring. Thus the normalised line intensities in jet lines and in stationary H$\alpha$ and He I emission more or less double in height over a period of a couple of days – this is readily visible in Fig. 2 of [9] – and this happens in the right place for the first three eclipses. Thereafter flaring makes the jet line intensities and flaring stationary lines less reliable for this technique, although the effect is certainly present, even in the red and blue flare components, (see Figs. 1, 4). The most reliable signal after day + 281 is in fact to be found in O I 8446 Å , a line that is strong and split and shows no traces of flaring effects. The O I 8446 Å spectra from the data set of [9] have not been published; a single example is shown in Fig. 2 of [14] (for day +274). I gave some discussion of these spectra (and their very different characteristics from those of O I 7772 Å) in [15]. I find that the data are adequate to conclude that the system eclipses were following the Goranskii ephemeris to within $\sim$ 0.03 orbital periods, 0.4 days.

The question then arising is how easily the supposed accretion disk lines could evade eclipse if the radius of the companion is in fact a more plausible 0.4$A$-0.5$A$. A simple extension of the argument used in section 4 of [5] shows that if the radius of the companion were $\sim$ 0.4$A$ then the red and blue extremes would be simultaneously visible if the observations on +294 and +307 differed from the Goranskii ephemeris by $\sim$ 0.03 orbital periods; if the radius were as great as 0.5 $A$ by $\sim$ 0.05. If the postulated thin disk abuts the photosphere, we have a better indicator. Represent the photosphere as a disk on the sky of radius $\sim$0.1$A$. It would not be eclipsed at all if $a/A \lesssim$ 0.1 . For $a/A \simeq$0.3 the interval between first and last contact in the eclipse is 1.5 days and full eclipse is reached. For an eclipse that only obscures half of the photospheric disk, $a/A \simeq$ 0.2, the interval between first and last contact is a single day. Both before and after the onset of the flare eclipses last roughly the same length of time and that time is over 2 days, see Fig.4. It is also the case that the spectral shape of the continuum exhibits no change during and after the onset of the flare (Fig.4). I regard these pieces of evidence as the strongest against the naïve interpretation of these data in [5]. The lack of eclipses of the extreme features and the duration of eclipses of the photosphere suggest that the H$\alpha$ flares do not originate in Keplerian orbits about the compact object. These aspects are discussed in more detail below.

**The photosphere of the accretion region.**

It was suggested in [5] that the appearance and disappearance of optical flaring might be associated with episodes of overflow from the Roche lobe of the Companion feeding the accretion disk via the $L_1$ point, or perhaps with some unspecified clouds breaking to reveal it. Such phenomena would be expected to affect the photosphere of the accretion region surrounding the compact object. The photosphere has a radius of very roughly 0.5 $A$ and is responsible for the continuum underlying the line spectra; if the line elements flaring in the visible and infrared are generated in a typical accretion disk they would have to be formed deep within the photosphere surrounding the accreting region. As more data have accumulated over the years, this has seemed increasingly unlikely. In the light of the results of [1] I have examined such data in [9] as exist on the state of the photosphere both before and after the flaring outburst commencing at approximately day + 287. In the absence of photometric data all that can be said about the eclipse episodes is that both before and after the outburst those data are consistent with eclipses lasting the same length of time (over 2 days between first and last contact) and cutting out about half of the continuum radiation from the photosphere (see Fig. 4). The eclipses last far too long for a photosphere of radius $a/A \approx 0.1$. The spectral shape of the continuum also does not change with the onset of flaring. (An example is to be found in Fig.3, taken from Fig. 2 of [13].)

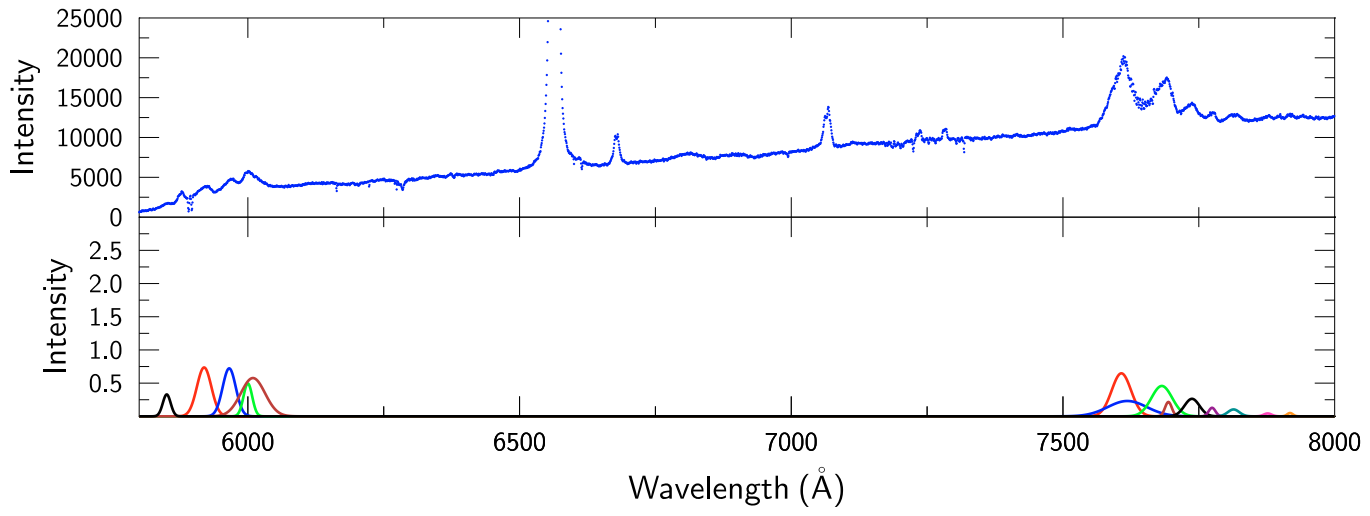

Fig.3 The spectrum of SS 433 on day +260. The stationary H$\alpha$ and He I lines are very clear, sitting on top of the sloping continuum, for which the ratio of 6500 Å to 8000 Å is 0.47. The lower panel shows extracted jet lines.

The continuum intensity rises rather linearly with wavelength, such that the intensity at 6500 Å divided by the intensity at 8000 Å is 0.47. Similar data for other days have not been published but I have access to them. Every day between + 245 and + 321 for which there are data shows a shape consistent with that same index, with the exception of two days (at different orbital and nutational phases), Fig.4.

**Evidence against emission line spectra from accretion disk orbits**

Material orbiting the compact object at $\sim 700\ km\ s^{-1}$ would have to be at a radius less than $\sim 0.1\ A$. This is deep inside the Roche lobe of the compact object and much less than the rough estimate for the photospheric radius of $\sim 0.5\ A$. In the absence of breaches in the photosphere, it hardly seems possible that H$\alpha$ line spectra are formed or can survive passage through an intact photosphere; a thin accretion disk radiating in H$\alpha$ would have to be feeding into the photosphere. There being no signs of disturbance of the photosphere during flaring in the line spectra, the implication is that the flare spectra are formed in a chromosphere external to the photosphere – a low density shell or a component of wind perhaps. The rotating and expanding shell proposed by GRAVITY [1] might be an example. Whatever the true nature of the source may be, there seems to be strong evidence that the line spectra are not formed deep within the photosphere. This to be weighed against the evidence in favour; that the pair of components to the red and blue track the phase of the compact object, just as radiation from a classic accretion disk would, if not immersed in a photosphere.

Radiation from an equatorial wind, launched at $\sim 1000\ km\ s^{-1}$ more or less in the plane of the orbit could potentially match the spectral observations, but would not provide the apparent rotation elements of GRAVITY. The problem remains: how could the material observed by GRAVITY pick up angular momentum about the binary system $\sim 1000A\ km^2\ s^{-1}$?

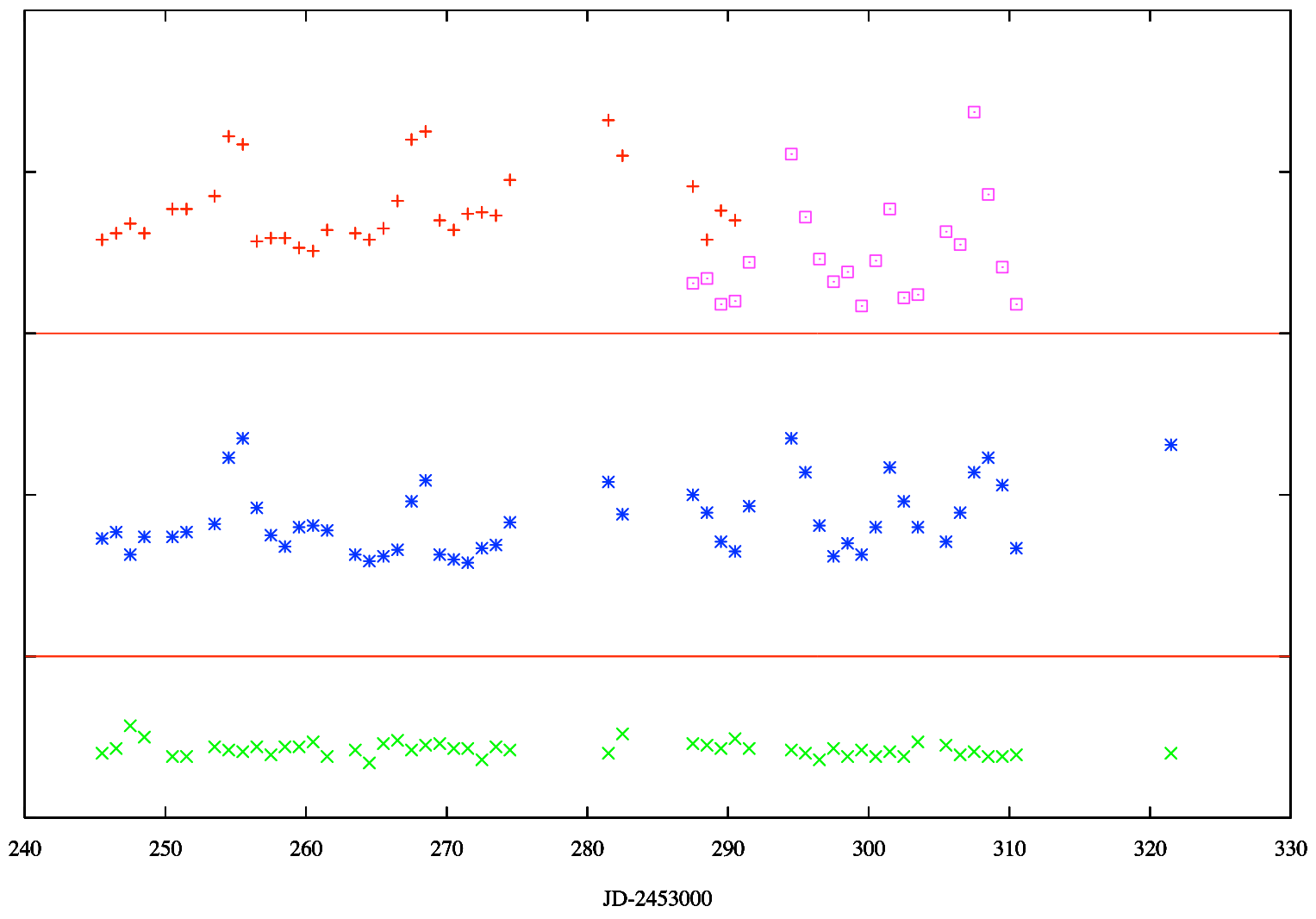

Fig.4 Assembled in this figure are the pieces of evidence weighing against the flares revealing a source deep within the photospheric radius. The lowest panel shows the ratio of continuum at 6500 Å to 8000 Å,×. There is no significant change after flaring commences at +287. The middle panel shows the ratio of O I 8446 Å to the local continuum, ∗. This ratio roughly doubles in every eclipse and is certainly not narrower after flaring is initiated. The photosphere is not blown away. The uppermost panel shows the ratio of He I 6678 Å to local continuum, +. This checks the eclipses against O I, until He I becomes unreliable. On the right of the uppermost panel is the ratio of the red flare component to background continuum,⊡. The two eclipses are clear, as is the effect of a period of obscuration.

Curiously, there does seem to be a single episode of photospheric obscuration. This is of significance for two reasons. First, this obscuration does not affect the extreme red and blue flare components – further evidence that they do not originate deep within the radius of the photosphere. Over the few days +300 to +304 inclusive the normalised intensities in both H$\alpha$ and O I 8446 Å vary in a way very similar to eclipse by the Companion, but these days are between two successive eclipses. The observed photospheric intensity appears to have dropped and recovered over this short period. This might be due to clouds drifting across the line of sight from the photosphere. Secondly, the jets vanished on day +302 and did not reappear until +306 [6, 13]. The possibility of a link between a rare cloud and switching off the jets is of interest.

## The source of the flares

Attributing the flare red and blue components of the H$\alpha$ and Br$\gamma$ to a source local to the compact object suffers the objection that it would seem not to be able to account for fast rotating material far out from the system, observed with GRAVITY, but the strong arguments related to the photosphere are valid even without GRAVITY. The strongest argument in favour is that the H$\alpha$ lines remember the orbital phase of the compact object quite well.

There is a potential explanation that can reconcile these conflicting elements and certain other issues. If the Companion fills or overfills its Roche lobe, material will spill out through the $L_2$ point, such a source sharing the orbital phase of the compact object and at a radius from the centre of mass of SS 433 of 1.22$A$. The speed of the $L_2$ point is 368, possibly as high as 400 $kms^{-1}$ and the specific angular momentum 422$A$, possibly as high as 487$A$ $km^2s^{-1}$. It seems safe to conclude that the asymptotic specific angular momentum of material leaving the $L_2$ point could be as much as 500$A$ $km^2s^{-1}$. (For comparison, the specific angular momentum of the compact object is 101$A$ $km^2s^{-1}$.)

$L_2$ ejection seems promising for explaining the GRAVITY results and can address the way the flare lines follow the motion of the compact object. Material leaving $L_2$ at speeds greater than the local escape velocity will have changed the orbital velocity after half an orbital period. Material ejected when viewed approaching is thus compared with material receding at a different velocity and on the opposite side of the system; the pattern cycles from orbit to orbit. This scenario was originally discussed in considerable detail in [16] and Fabrika's prescient paper dealing specifically with SS 433, [17]. A more recent paper is [18]. Further, the features of several other data sets fit well into this picture (see Appendix) and the intermittent flaring might be a consequence of mass ejections from the Companion to flood the outer Roche lobe. The variation of the mean velocity of course depends on details. (An example is discussed in [19].)

## Conclusions

It may well be that the fast outbreaks observed in flaring do not correspond to glimpses of the rim of a classic accretion disk, despite flare memories of the motion of the compact object; I now think it likely that these outbreaks have the same origin as the ejected material reported by GRAVITY and that expelled though $L_2$. It is not however clear to me whether the GRAVITY observations could be consistent with the specific angular momentum $\sim 500A\ km^2s^{-1}$rather than $\sim 1000A\ km^2s^{-1}$ . If not, it would be nice to know how this material acquired its additional high angular momentum about the binary system of SS 433.

## Appendix

Here I discuss briefly three other data sets that are relevant. The first consists of a number of spectra taken during a single orbit, in the infrared [7], and shows features very similar to the latter period in [9], as analysed in [12]. The Br$\gamma$ line is complex; analysed as a superposition of Gaussian profiles it reveals high speed wind and two pairs of much narrower components. One corresponds to the circumbinary disk with an orbital speed of $\sim$ 250 km $s^{-1}$, the pair of more extreme components are offset by $\sim\pm600\ km\ s^{-1}$. This pair was interpreted in [7] as from the accretion disk of the compact object and it was suggested that perhaps it was visible in the infrared but only intermittently in the red. However, the ten or so spectra spread over a complete orbit show no memory whatsoever of the motion of the compact object. On the other hand, the extreme blue component is absent in the single spectrum taken marginally before eclipse.

The second set of spectra relevant to these notes consists of five sets taken in the H and K bands, intermittently in 2014 and 2015. The relevant material is displayed in Figs. 1 and 7 of [10], profiles of emission lines in the Brackett sequence. They are mentioned here because these lines are much narrower than the Br$\gamma$ sequence in [7]. The latter required fast components; the Br sequence in [10] do not contain such fast components [12] but are readily explained by lines from a wind and lines from the circumbinary disk, just like the H$\alpha$ lines in the first part of the sequence in [9], see also [12]. Even in the far infrared the hydrogen emission lines appear only intermittently.

Finally, a sequence of parasitic observations covering the Balmer H$\alpha$ and H$\beta$ emission lines covered a large part of a precession period intermittently. No more than a few spectra were taken during any single orbit. Throughout this sequence both H$\alpha$ and H$\beta$ displayed high speed components to both the red and the blue. The displacements varied from $\sim 500\ km\ s^{-1}$ to $\sim 1000\ km\ s^{-1}$, Fig.3 of [7]. There is little prospect of being able to detect any systematic variation with the orbital phase, nor indeed with precessional phase. For present purposes, the significance of these observations is that the displacements to red and to blue can be far from symmetric and can vary substantially from one spectrum to the next, over periods of a few days. Whatever the source may be, it does not have the stability one would look for were it to be a Kepler accretion disk, unless clouds came and went.

## References

[1] I. Waisberg et al, A&A **623**, A47 (2019)
[2] M. G. Bowler, A&A **619**, L4 (2018)
[3] A. M. Cherepashchuk et al, MNRAS **479**, 4844 (2018)
[4] M. A. Dopita & A. M. Cherepashchuk, Vistas in Astronomy **25**, 51 (1981)
[5] M. G. Bowler, A&A **516**, A24 (2010)
[6] K. M. Blundell et al, MNRAS **417**, 2401 (2011)
[7] S. Perez M.& K. M. Blundell MNRAS **408**, 2 (2010)
[8] M. G. Bowler, arXiv:2009.00589
[9] L. Schmidtobreick & K. Blundell PoS (MQW6) 094 (2006)
[10] E.L. Robinson et al, **Ap. J.** 841, 79 (2017)
[11] V. P. Goranskii et al, Astron. Rep. **41**, 656 (1996)
[12] M. G. Bowler, arXiv: 1708.02555 (2017)
[13] K. M. Blundell et al, A&A **474**, 903 (2007)
[14] L. Schmidtobreick & K. Blundell, Astrophys. Space Sci. **304**, 271 (2006)
[15] M. G. Bowler, arXiv:1512.06885v2 (2021)
[16] F. H. Shu et al, Ap.J. **229**, 223 (1979)
[17] S. N. Fabrika, MNRAS **261**, 241 (1993)
[18] O. Pejcha et al, MNRAS **455**, 4351 (2016)
[19] M. G. Bowler, A&A **521**, A81 (2010)